\documentclass[Physsubmission, Phys]{SciPost}

\hypersetup{
    colorlinks,
    linkcolor={red!50!black},
    citecolor={blue!50!black},
    urlcolor={blue!80!black}
}

\usepackage[bitstream-charter]{mathdesign}
\DeclareSymbolFont{usualmathcal}{OMS}{cmsy}{m}{n}
\DeclareSymbolFontAlphabet{\mathcal}{usualmathcal}

\newcommand{\sm}{\hbox{$\bigcirc$\kern-0.6em\hbox{s} }}

\newcommand{\NN}{\hbox{I\kern-.2em\hbox{N}}}  
\newcommand{\ZZ}{{{\rm Z}\kern-.28em{\rm Z}}} 
\newcommand{\RR}{\mathop{{\rm I}\kern-.2em{\rm R}}\nolimits} 
\newcommand{\RRe}{\mathop{{\rm I}\kern-.2em{\rm Re}}\nolimits} 
\newcommand{\QQ}{\hbox{l\kern-.36em\hbox{Q}}}  
\newcommand{\CC}{\hbox{{\textsf I}\kern-.47em\hbox{C}}}
\newcommand{\nop}{\hbox{{\textsf I}\kern-.47em\hbox{O}}}

\def\TREV{{{}^\triangleleft\kern-1.5pt\texttt{T}}}
\def\trev{{{}^\triangleleft\kern-3.2pt\texttt{t}}}
\def\SREV{{{}_\triangleleft\kern-2pt\texttt{S}}}
\def\srev{{{}_\triangleleft\kern-2.2pt\texttt{s}}}
\newcommand{\Id}{\hbox{\sl 1\kern-0.25em\hbox{I}}}
\definecolor{darkgreen}{rgb}{0,0.5,0}
\definecolor{violet}{rgb}{0.4,0,0.5}
\definecolor{dblue}{rgb}{0,0,0.7}
\definecolor{dred}{rgb}{0.7,0,0}
\definecolor{grey}{rgb}{0.5,0.5,0.5}
\definecolor{white}{rgb}{1,1,1}

\begin{document}

\begin{center}{\Large \textbf{
Group theoretical derivation of consistent particle theories\\
}}\end{center}

\begin{center}
Giuseppe Nistic\`o\textsuperscript{1},

\end{center}

\begin{center}
{\bf 1} Universit\`a della Calabria, Italy
\\
{\bf 2} INFN, gr. collegato di Cosenza, Italy
\\
* giuseppe.nistico@unical.it
\end{center}

\begin{center}
\today
\end{center}


\definecolor{palegray}{gray}{0.95}
\begin{center}
\colorbox{palegray}{
  \begin{tabular}{rr}
  \begin{minipage}{0.1\textwidth}
    \includegraphics[width=20mm]{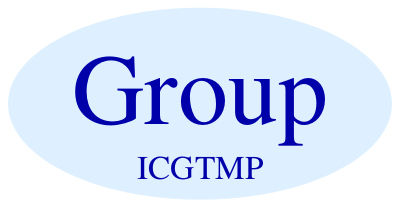}
  \end{minipage}
  &
  \begin{minipage}{0.85\textwidth}
    \begin{center}
    {\it 34th International Colloquium on Group Theoretical Methods in Physics}\\
    {\it Strasbourg, 18-22 July 2022} \\
    \doi{10.21468/SciPostPhysProc.?}\\
    \end{center}
  \end{minipage}
\end{tabular}
}
\end{center}

\section*{Abstract}
{\bf
Current quantum theories of an elementary free particle assume unitary space inversion and anti-unitary
time reversal operators. In so doing robust classes of possible theories are discarded.
The present work shows that consistent theories can be derived through a strictly deductive
development from the principle of relativistic invariance and position covariance, also with
anti-unitary space inversion and unitary time reversal operators. In doing so the class of possible 
consistent theories is extended for positive but also zero mass particles.
In particular, consistent theories for a Klein-Gordon particle are derived and the non-localizability theorem
for a non zero helicity massless particle is extended.
}

\vspace{10pt}
\noindent\rule{\textwidth}{1pt}
\tableofcontents\thispagestyle{fancy}
\noindent\rule{\textwidth}{1pt}
\vspace{10pt}
\section{Introduction}
Relativistic quantum theories of single free particle can be deductively derived from the 
principles of {\sl relativistic invariance} and {\sl covariance} \cite{Wigner1} -\cite{Wightman}; 
the first principle implies that the Hilbert space of the quantum theory of a free particle
must admit a {\sl transformer triplet} $(U,\SREV,\TREV)$ formed by a unitary representation $U$ of the universal covering
group $\tilde{\mathcal P}_+^\uparrow$ of the proper orthochronus Poincar\'e group ${\mathcal P}_+^\uparrow$
and by the operators $\SREV$ and $\TREV$, which realize
the quantum transformations implied by the transformations of ${\mathcal P}_+^\uparrow$, by space inversion $\srev$ 
and by time reversal $\trev$, respectively.
Yet the literature, 
except some works \cite{DeberghAl}\cite{Varlamov} with specific aims different from the present one,
excludes transformer triplets with $\SREV$ anti-unitary or with $\TREV$ unitary,
from the pionering works of
Wigner, Bargmann \cite{Wigner1} -\cite{Bargmann1},
to subsequent investigations \cite{Wightman} -\cite{Jordan1} .
In so doing robust classes of triplets, and hence of possible theories, are lost. For instance, 
there is no such a triplet for a consistent theory of Klein-Gordon 
particles\footnote{Klein-Gordon theory, indeed, was obtained through canonical quantization \cite{Klein},\cite{Gordon},
but it predicts inconsistencies, such as negative probabilities\cite{Nistico1}.}.
\par
The motivation for the exclusion of $\TREV$ unitary or $\SREV$ anti-unitary was their implication of negative spectral
values for the hamiltonian operator $P_0$, values deemed inconsistent because $P_0$ was identified with the {\sl positive}
relativistic kinetic energy operator  $E_{kin}=\mu(1-{\dot{\bf Q}}^2)^{-1/2}$,
where $\dot{\bf Q}$ is the ``velocity'' operator.
But remark 3.1 shall show that the hamiltonian operator $P_0$ does not always coincide with $E_{kin}$, 
so that a unitary $\TREV$ or an anti-unitary $\SREV$ can be consistent.
\par
In the present article we show how a strictly deductive development of consistent quantum theories of elementary free particle can be successfully carried out without
{\it apriori} preclusions about the unitary or anti-unitary character of $\SREV$
or $\TREV$.
As results, classes of consistent possible theories for a positive mass particle
are expicitly identified, which meaningfully extend the class of the current theories; in particular, 
consistent theories of Klein-Gordon particle are derived.
Also in the case of a massless particle the approach extends the class of possible theories.
Furthermore, the non-localizability theorem for non zero helicity massless particles is extended to the new theories 
with $\TREV$ unitary or $\SREV$ anti-unitary.
\vskip.5pc
Section 2 shows how the relativistic invariance principle implies that every theory of elementary free particle admits a
transformer triplet.
In section 3 the class of possible consistent theories for a positive mass particle is identified;
this class contains consistent theories with $\SREV$ anti-unitary, e.g. consistent theories of Klein-Gordon particle.
Section 4 identifies the class of consistent theories for a zero mass elementary free particle;
once again, besides the current theories, it contains theories with $\SREV$ anti-unitary or $\TREV$ unitary.
A more accurate and more general argument is presented, 
which denies localizability of non zero helicity mass zero particles

\section{General implications of Poincar\'e invariance}
\subsection{Prerequisites and notation}
First of all, it is worth to fix the notaion for any quantum theory based on a Hilbert space $\mathcal H$:
\begin{description}
\item{-}
$\Omega({\mathcal H})$ denotes the set of all self-adjoint operators representing observables;
\item{-}
${\mathcal S}({\mathcal H})$ denotes the set of all density operators $\rho$ identified with quantum states;
\item{-}
${\mathcal U}({\mathcal H})$ denotes the group  of all unitary unitary operators;
\item{-}
${\mathcal V}({\mathcal H})$ is the larger group of all unitary or anti-unitary operators.
\end{description}
The Poincar\'e group $\mathcal P$ is a very important mathematical structure for the present work, 
because it is the group of symmetry transformations for a free particle.
$\mathcal P$ is the group generated by ${\mathcal P}_+^\uparrow\cup\{\trev,\srev\}$, where
${\mathcal P}_+^\uparrow$ is the proper orthochronus Poincar\'e group, $\trev$ and $\srev$ 
are the time reversal   and space inversion  transformations.
The proper orthochronus group
${\mathcal P}_+^\uparrow$ is a connected group generated by 10 one-parameter subgroups, namely the subgroup
${\mathcal T}_0$ of time translations, the three subgroups
${\mathcal T}_j$ ($j=1,2,3$) of  spatial translations, the three subgroups
${\mathcal R}_j$  of spatial rotations, the three subgroups
${\mathcal B}_j$ of Lorentz boosts, relative to the three spatial axes $x_j$.
Time reversal $\trev$ and space inversion $\srev$ are not connected with the identity transformation $e\in\mathcal P$.
Given any vector $\underline x=(x_0,{\bf x})\in\RR^4$, where $x_0$ is called the {\sl time component}
of $\underline x$ and ${\bf x}=(x_1,x_2,x_3)$ is called the {\sl spatial component} of $\underline x$,
time reversal $\trev$ transforms $\underline x=(x_0,{\bf x})$ into $(-x_0,{\bf x})$ and space inversion $\srev$
transforms $\underline x=(x_0,{\bf x})$ into $(x_0,-{\bf x})$.
\par
The universal covering group of ${\mathcal P}_+^\uparrow$ is the
semidirect product $\tilde{\mathcal P}_+^\uparrow=\RR^4\sm  SL(2,\CC)$
of the time-space translation group $\RR^4$ and the group
$SL(2,\CC)=\{\underline\Lambda\in GL(2,\CC)\mid \det\underline\Lambda=1\}$.
Accordingly, $\tilde{\mathcal P}_+^\uparrow$ is simply connected and
there is a canonical homomorphism
${\textsf h}:\tilde{\mathcal P}_+^\uparrow\to{\mathcal P}_+^\uparrow$,
$\tilde g\to {\textsf h}(\tilde g)\in{\mathcal P}_+^\uparrow$, which restricts to an isomorphism within a small
enough neighborhood of the identity
$(0,\Id_{{_{\CC^2}}})$ of $\tilde{\mathcal P}_+^\uparrow$.
By $\tilde{\mathcal T}_0$, $\tilde{\mathcal T}_j,\tilde{\mathcal R}_j$, $\tilde{\mathcal B}_j$,
$\tilde{\mathcal L}_+^\uparrow$ we denote the subgroups of $\tilde{\mathcal P}_+^\uparrow$ which
correspond to the subgoroups ${\mathcal T}_0$, ${\mathcal T}_j,{\mathcal R}_j$, ${\mathcal B}_j$,
${\mathcal L}_+^\uparrow$  of ${\mathcal P}_+^\uparrow$,
through the homomorphism $\textsf h$.

\subsection{Quantum theoretical implications for an elementary free particle}
Since a free particle is a particular kind of isolated system, we begin by showing the derivation
of the general structure of the quantum theory of an isolated system.
By $\mathcal F$ we denote the class  of the (inertial) reference frames that move uniformly with respect to each other.
A physical system is an {\sl isolated system} if the following {\sl invariance principle} holds.
\begin{description}
\item[$\mathcal{IP}$]{\sl
The theory of an isolated system is invariant with respect to changes of frames within $\mathcal F$.}
\end{description}
If $\Sigma$ belongs to $\mathcal F$, then  $\Sigma_g$ denotes the frame related to $\Sigma$ by such $g$,
for every $g\in\mathcal P$.
Given an observable $\mathcal A$ represented by the operator $A\in\Omega({\mathcal H})$, let ${\mathcal M}_A$ be a procedure to measure $\mathcal A$;
then the invariance principle implies that another measuring procedure ${\mathcal M}_A'$ must exist, which is with respect to $\Sigma_g$ identical to what is ${\mathcal M}_A$
with respect to $\Sigma$, otherwise the principle $\mathcal{IP}$ would be violated. Hence, $\mathcal{IP}$ implies the existence \cite{Nistico1} of the so called {\sl quantum transformation associated to $g$}, i.e.,  of a mapping
$$
S_g:\Omega({\mathcal H})\to\Omega({\mathcal H})\,,\quad A\to S_g[A]\,,
$$
where ${\it S_g}[A]$ is the self-adjoint operator that represents the observable measured by ${\mathcal M}_A'$.
\par
To every element $\tilde g$ of the covering group $\tilde{\mathcal P}_+^\uparrow$ we can associate the quantum transformation
$S_{\textsf h(\tilde g)}\equiv S_{\tilde g}$  through the canonical homomorphism $\textsf h$.
In \cite{Nistico1} it is proved that the properties of quantum transformations,
under a continuity condition for $\tilde g\to S_{\tilde g}$, imply that
\begin{description}
\item{\it Imp.1.}
a continuous unitary
representation $U$ of $\tilde{\mathcal P}_+^\uparrow$ exists such that
$S_{\tilde g}[A]= U_{\tilde g}A U_{\tilde g}^{-1}$, and
\item{\it Imp.2.}
two operators $\SREV,\TREV\in{\mathcal V}({\mathcal H})$ exist such that
$S_\srev[A]=\SREV A\SREV^{-1}$ and $S_\trev[A]=\TREV A\TREV^{-1}$.
\end{description}
Thus,
the principle $\mathcal{IP}$ has the following fundamental implication.
\begin{description}
\item[($\textsf{FI}$)]{\sl The quantum theory of an isolated system admits a transformer triplet $(U,\SREV,\TREV)$ such that implications {\it Imp.1} and {\it Imp.2} hold.}
\end{description}
Given a transformer triplet $(U,\SREV,\TREV)$, let $P_0, P_j, J_j, K_j\in\Omega({\mathcal H)}$ be the 
{\sl selfadjoint generators} of $U$; so \cite{Nistico1},
if $\tilde g\in\tilde{\mathcal T}_0$ (resp., $\tilde{\mathcal T}_j$, $\tilde{\mathcal R}_j$, $\tilde{\mathcal B}_j$) is identified by the parameter $t$ (resp., $a$,
$\theta$, $u$), then
$$U_{\tilde g}=e^{iP_0t},\quad(\hbox{resp., }U_{\tilde g}=
e^{iP_ja},\;U_{\tilde g}=e^{J_j\theta}, \;U_{\tilde g}=e^{iK_j\frac{1}{2}\ln\frac{1+u}{1-u}}).\eqno(1)$$
The generator $P_0$ relative to time translations is the {\sl hamiltonian operator}, so that
$$
(i) \quad\frac{d}{dt}A_t\equiv \dot A_t=i[P_0,A_t], \qquad (ii)\quad \frac{d}{dt}\rho_t\equiv\dot\rho_t=-i[P_0,\rho_t].\eqno(2)
$$
By {\sl ``elementary'' free particle} we mean an isolated system whose quantum theory has
a {\sl unique} three-operator ${\bf Q}\equiv(Q_1,Q_2,Q_3)$ with $Q_j\in\Omega({\mathcal H})$, 
called {\sl position} operator, such that
$(U(\tilde{\mathcal P}_+^\uparrow),\SREV,\TREV; {\bf Q})$ is an {\sl irreducible} system of operators, and satisfying the following conditions.
\begin{itemize}
\item[({\it Q}.1)]
$\;[Q_j,Q_k]=\nop$, for all $j,k=1,2,3$;
this condition establishes that a measurement of position yields
all three values of the coordinates of the same specimen of the system.
\item[({\it Q}.2)]
$\;$ For every $g\in\mathcal P$,
the position operator ${\bf Q}$ and the transformed position operator
$S_g[{\bf Q}]$ satisfy
the transformation properties of position with respect to $g$.
\end{itemize}
\vskip.5pc
As proved in \cite{Nistico1}, the transformer triplet $(U,\SREV,\TREV)$ of the quantum theory of
an elementary free particle must be {\sl irreducible}. Thus,
the identification of all possible theories of an elementary free particle can be carried out in two steps:
first by identifying all irreducible transformer triplets $(U,\SREV,\TREV)$, 
and then selecting those triplets for which a unique position operator $\bf Q$ exists.
\vskip.8pc
The mathematical group structural properties of $\mathcal P$ imply 
\cite{Nistico1},\cite{Nistico2} that each irreducible triplet  $(U,\SREV,\TREV)$ is
characterized by a number $\mu\in\CC$, called {\sl mass}, with $\mu^2\in\RR$, such that $P_0^2-{\bf P}^2=\mu^2\Id$.
\section{Quantum theories of positive mass elementary free particle}
To identify the positive mass possible theories, we shall identify the irreducible triplets with $\mu>0$; then, the triplets admitting a three-operator $\bf Q$ satisfying ({\it Q}.1), ({\it Q}.2) are singled out.
\subsection{Positive mass irreducible triplets}
Following \cite{Nistico1},
for any pair $(\mu,s)$, where $\mu>0$ and $s$ is an
integral or half-integral number $s\in\frac{1}{2}\NN$ called {\sl spin},  there is at least one irreducible triplet.
Conversely, every irreducible triplet is characterized by one such a pair.
The following theorem yields a first classification. 
\vskip.5pc
{\bf Theorem 3.1.}
{\sl If $(U,\SREV,\TREV)$ is an irreducible triplet with non-negative mass $\mu\geq 0$, then
\par\noindent
i)
$\sigma(P_0)=(-\infty,-\mu]$
or $\sigma(P_0)=[\mu,\infty)$
or $\sigma(P_0)=(-\infty,-\mu]\cup[\mu,\infty)$,
where $\sigma(P_0)$ is the spectrum of $P_0$. 
\par
Moreover, 
$\sigma(P_0)=(-\infty,-\mu]\cup[\mu,\infty)$ if and only is $\TREV$ is unitary or $\SREV$ is anti-unitary.
\par\noindent
ii) 
Each class  ${\mathcal I}(\mu,s)$ of all irreducible triplets with positive mass $\mu>0$ decomposes as
$$
{\mathcal I}(\mu,s)={\mathcal I}^-(\mu,s)\cup {\mathcal I}^+(\mu,s)\cup{\mathcal I}^{-+}(\mu,s).\eqno(3)
$$
where
${\mathcal I}^-(\mu,s)$, ${\mathcal I}^{+}(\mu,s)$ and ${\mathcal I}^{-+}(\mu,s)$ are respectively the classes 
of irreducible triplets with $\sigma(P_0)=(-\infty,-\mu]$, $\sigma(P_0)=[\mu,\infty)$ and
$\sigma(P_0)=(-\infty,-\mu]\cup[\mu,\infty)$.}
\vskip.5pc\noindent
The representation $U$ of a triplet in ${\mathcal I}^+(\mu,s)$ or ${\mathcal I}^-(\mu,s)$
can be irreducible or not. 
We refer to \cite{Nistico1} for a complete identification of the irreducible triplets of ${\mathcal I}^\pm(\mu,s)$ 
with $U$ irreducible. Therein also instances of triplet in ${\mathcal I}^+(\mu,s)$ and 
${\mathcal I}^+(\mu,s)$ with $U$ reducible are explicitly shown.
\vskip.5pc
The representation $U$ of a triplet in ${\mathcal I}^{-+}(\mu,s)$ is always reducible \cite{Nistico1}, namely
$U=U^+\oplus U^-$ where $U^\pm$ belongs to a triplet in ${\mathcal I}^\pm(\mu,s)$. Moreover,
$U^+$ is reducible if and only if $U^-$ is reducible. 
\par
The class of all irreducible triplets of ${\mathcal I}^{-+}(\mu,s)$ with $U^+$ irreducible can be found in \cite{Nistico1},
where also triplets of ${\mathcal I}^{-+}(\mu,s)$ with $U^+$ reducible are concretely shown.
\subsection{Theories of elementary free particle with positive mass}
To determine the possible theories of positive mass elemetary free particle, 
we have to select irreducible triplets of ${\mathcal I}(\mu,s)$ identified in \cite{Nistico1}
for which  a position $\bf Q$ satisfying ({\it Q}.1) and ({\it Q}.2) exists. Condition ({\it Q}.2) can be only partially imposed. In fact, while the covariance properties with respect to translations, rotations, time reversal and space inversion are known and explicitly expressed by the following relations\cite{Nistico1}
$$
(i)\;\;[Q_j,P_k]=i\delta_{jk},\quad(ii)\;\;[J_j,Q_k]=i\epsilon_{jkl}Q_l,
\quad(iii)\;\;\TREV{\bf Q}={\bf Q}\TREV,\quad(iv)\;\;\SREV{\bf Q}=-{\bf Q}\TREV,
\eqno(4)
$$
the explicit relations that establish the transformation properties of position with respect 
to boosts are not available, yet \cite{Nistico1}. However, conditions (4) are sufficient to uniquely identify $\bf Q$
for some subclasses of irreducible triplets, according to the following theorem  \cite{Nistico1}.
\vskip.5pc
{\bf Theorem 3.2.} 
{\sl Given a triplet in ${\mathcal I}^+(\mu,0)$ with $U$ irreducible
there is a unique three-operator $\;\bf Q$, satisfying ({\it Q}.1) and (4). Modulo unitary isomorphism, the resulting theory  has Hilbert space ${\mathcal H}=L_2(\RR^3,\CC^{2s+1},d\nu)$, where
$d\nu({\bf p})=\frac{dp_1dp_2dp_3}{p_0}$ with $p_0=\sqrt{\mu^2+{\bf p}^2}$, 
\begin{description}
\item[--]
generators defined by
$(P_j\psi)({\bf p})=p_j\psi({\bf p})$,
$(P_0\psi)(\underbar p)= p_0\psi(\underbar p)$, $J_k={\textsf J}_k^{(0)}$, $K_j={\textsf K}_j^{(0)}$,\hfill{$(5)$}
\item[$\;$]
where
${\textsf J}_k^{(0)}=-i\left(p_l\frac{\partial}{\partial p_j}-p_j\frac{\partial}{\partial p_l}\right)$,\quad
${\textsf K}_j^{(0)}=ip_0\frac{\partial}{\partial p_j}$;

\item[--] 
$\SREV=\Upsilon$,\quad $\TREV={\mathcal K}\Upsilon$, where 
${\mathcal K}$ and
$\Upsilon$ are defined by
${\mathcal K}\psi({\bf p})=\overline{\psi({\bf p})}$,
$(\Upsilon\psi)({\bf p})=\psi(-{\bf p})$.
\item[--]
The position operator is
${\bf Q}={\bf F}$, where $\bf F$ is the Newton-Wigner \cite{Wigner3} operator defined by 
$$
F_j=i\frac{\partial}{\partial p_j}-\frac{i}{2p_0^2}p_j.\eqno(6)
$$
\end{description}
Analogously, there is only one theory based on a triplet in $\;{\mathcal I}^-(\mu,0)$ with $U$ irreducible. 
It differs from that in
$\;{\mathcal I}^+(\mu,0)$ by
$P_0=-p_0$ and $K_j=-{\textsf K}_j^{(0)}$.
\vskip.5pc
There are only two theories based on triplets of ${\mathcal I}^{-+}(\mu,0)$ with $U^+$ irreducible. They share the
Hilbert space and generators\footnote{
If $\psi\in L_2(\RR^3,\CC^{2s+1},d\nu)\oplus L_2(\RR^3,\CC^{2s+1},d\nu)$, we write $\psi\equiv \psi_1\oplus\psi_2\equiv\left[\begin{array}{c}\psi_1\cr \psi_2\end{array}\right]$,
$\psi_1,\psi_2\in  L_2(\RR^3,\CC^{2s+1},d\nu))$.}:
${\mathcal H}=L_2(\RR^3,\CC^{2s+1},d\nu)\oplus L_2(\RR^3,\CC^{2s+1},d\nu)$
\vskip.5pc\noindent
$P_j=\left[\begin{array}{cc}p_j&0\\ 0&p_j\end{array}\right],\quad 
P_0=\left[\begin{array}{cc}p_0&0\\ 0&-p_0\end{array}\right],\quad
J_k=\left[\begin{array}{cc}{\textsf J}^{(0)}_k&0\\ 0&{\textsf J}^{(0)}_k\end{array}\right],\quad
K_j=\left[\begin{array}{cc}{\textsf K}^{(0)}_j&0\\ 0&-{\textsf K}^{(0)}_j\end{array}\right]$.\hfill{(7)}\vskip1pc\noindent
The two theories differ for the different pairs $(\SREV_1,\TREV_1)$, $(\SREV_2,\TREV_2)$ of space inversion and time reversal operators; indeed 
$\;\SREV_1=\SREV_2=\left[\begin{array}{cc}0&1\\ 1&0\end{array}\right]{\mathcal K}$
while $\TREV_1={\mathcal K}\Upsilon\left[\begin{array}{cc}1&0\\ 0&1\end{array}\right]$
and
$\TREV_2=\left[\begin{array}{cc}0&1\\ 1&0\end{array}\right]$.
\par\noindent
For both theories the position operator is
 $\;{\bf Q}=\left[\begin{array}{cc}{\bf F}&0\cr 0&{\bf F}\end{array}\right]$.
}
 \vskip.8pc\noindent
For all triplets with $s>0$ ({\it Q}.1) and (4) are not sufficient \cite{Nistico1} to completely identify $\bf Q$.
\vskip.8pc\noindent
{\bf Remark 3.1.}
In both theories based on ${\mathcal I}^{-+}(\mu,0)$ the hamiltonian operator $P_0$ has also negative spectral values. 
But since the ``velocity'' is
$\dot{\bf Q}=\frac{d}{dt}{\bf Q}=i[P_0,{\bf Q}]=\left[\begin{array}{cc}\frac{\bf p}{p_0}&0\\0&-\frac{\bf p}{p_0}\end{array}\right]$, 
we compute that $E_{kin}=\mu(1-{\dot{\bf Q}}^2)^{-1/2}=p_0>\nop$, i.e. the theories are consistent.
\subsection{Conclusions for the positive mass case}
According to section 3.2, 
four classes of possible consistent theories are completely determined by following the present approach, with $U$ or $U^+$ irreducible.
However, the class of theories based on ${\mathcal I}^{\pm}(\mu,0)$ with $U$ reducible and the class of theories based on 
${\mathcal I}^{-+}(\mu,0)$ with $U^+$ reducible are not empty; concrete examples are given in \cite{Nistico1}. They are
{\sl new species theories}, i.e. they correspond to none of the known theories.
Hence, our approach extends the class of consistent theories of positive mass elementary spin 0 free particle.
\par
Moreover, it provides consistent theories for Klein-Gordon particles.
Indeed, by means of a unitary transformation,
operated by the operator $Z=Z_1Z_2$, where $Z_2=\frac{1}{\sqrt{p_0}}\Id$ and $Z_1$ is the inverse of the 
{\sl Fourier-Plancherel} operator, the theories based on ${\mathcal I}^{-+}(\mu,0)$ turn out to be
equivalent to two theories with Hilbert space
$\hat{\mathcal H}=Z\left(L_2(\RR^3,d\nu)\oplus L_2(\RR^3,d\nu)\right)\equiv L_2(\RR^3)\oplus L_2(\RR^3)$, 
with the self-adjoint generators \quad
${\hat P}_j=\left[\begin{array}{cc}-i\frac{\partial}{\partial x_j}&0\cr 0&-i\frac{\partial}{\partial x_j}\end{array}\right]$,\quad
${\hat P}_0=\sqrt{\mu^2-\nabla^2}\left[\begin{array}{cc}1&0\cr 0&-1\end{array}\right]$,
\par
${\hat J}_j=-i\left(x_k\frac{\partial}{\partial x_l}-x_l\frac{\partial}{\partial x_k}\right)\left[\begin{array}{cc}1&0\cr 0& 1\end{array}\right]$,\quad
${\hat K}_j=\frac{1}{2}\left(x_j\sqrt{\mu^2-\nabla^2}+\sqrt{\mu^2-\nabla^2}x_j\right)\left[\begin{array}{cc}1&0\cr 0&-1\end{array}\right]$,\par\noindent
while $\hat\TREV_1={\mathcal K}\left[\begin{array}{cc}1&0\cr 0&1\end{array}\right]$ , $\hat\SREV_1={\mathcal K}\Upsilon\left[\begin{array}{cc}0&1\cr 1&0\end{array}\right]$, and
$\hat\TREV_2=\left[\begin{array}{cc}0&1\cr 1&0\end{array}\right]$ and $\hat\SREV_2={\mathcal K}\Upsilon\left[\begin{array}{cc}0&1\cr 1&0\end{array}\right]$.
The position operator is
${\hat Q}_j=\left[\begin{array}{cc}x_j&0\cr 0&x_j\end{array}\right]$.
\par
These ${\hat P}_j,\;{\hat J}_j,\;\;{\hat K}_j,\;\hat{\mathcal H}$ are generators and Hilbert space of Klein-Gordon theory 
of spin-0 particle \cite{Klein}\cite{Gordon}.
However, since the position operator is the multiplication operator, the position probability density must be 
$\rho(t,{\bf x})=\vert \hat \psi_1(t,{\bf x})\vert^2+\vert \hat \psi_2(t,{\bf x})\vert^2$, hence non-negative. Thus, 
our extended class includes consistent theories  for Klein-Gordon 
particle free from the inconsistent negative probabilities of the early theory.
\vskip.5pc
It turns out \cite{Nistico1} that in all triplets with non zero spin, the position operator $\bf Q$ 
is not uniquely determined by ({\it Q}.1) and (4).
On the other hand, the transformation properties of position with respect to boosts,
expressed for instance by a relation for $[K_j,Q_k]$, are not available in order to better identify $\bf Q$ by imposing them.
\par
To each solution $\bf Q$ of ({\it Q}.1) and (4) there correspond a different $[K_j,Q_k]$, in general.
For instance, Dirac theory for spin 1/2 particle \cite{Dirac} is {\sl completely} characterized by the relation
$[K_j,Q_k]=-\frac{i}{2}(Q_j\dot Q_k+\dot Q_k Q_j)$ satisfied by the posistion operator of Dirac theory;
however, other solutions $\bf Q$ yielding other relations for $[K_j,Q_k]$ are theoretically consistent too.
\section{Quantum theories of zero mass elementary free particle}
Analogously to the positive mass case, the possible quantum theories of zero mass particle 
are determined first by identifying 
the class ${\mathcal I}_0$ of the irreducible 
transformer triplets with $\mu=0$, and then by selecting those triplets that admit a unique position operator.
According to theorem 3.1.i the class
${\mathcal I}_0$ decomposes as
${\mathcal I}^+_0={\mathcal I}^+_0\cup {\mathcal I}^-_0\cup{\mathcal I}^{-+}_0$,
where ${\mathcal I}^-_0$ (resp., ${\mathcal I}^+_0$, ${\mathcal I}^{-+}_0$) denotes the class
of irreducible triplets with $\sigma(P_0)=(-\infty,0]$ (resp., $\sigma(P_0)=[0,\infty)$, $\;\sigma(P_0)=\RR$).
\subsection{Zero mass irreducible triplets}
In \cite{Nistico2} the irreducible triplets of ${\mathcal I}^+_0$ and ${\mathcal I}^-_0$ with $U$
irreducible, and of  ${\mathcal I}^{-+}_0$ with $U^+$ irreducible are completely identified. The results are 
collected by the following statement.
\vskip.5pc
{\bf Theorem 4.1.}
{\sl Modulo unitary isomorphisms, there is only one triplet $(U,\SREV,\TREV)$ in ${\mathcal I}^+_0$ 
and in ${\mathcal I}^-_0$ with $U$ irreducible,
whose Hilbert space is ${\mathcal H}=L_2(\RR^3,d\nu)$, and
\par
$(P_j\psi)({\bf p})=p_j\psi({\bf p})$, \quad$P_0\psi({\bf p})=\pm p_0\psi({\bf p})$,\quad
$J_j={\textsf J}_j^{(0)}$,\quad $K_j=\pm {\textsf K}_j^{(0)}$, \quad $\SREV=\Upsilon$,\quad $\TREV={\mathcal K}\Upsilon$.
\vskip.5pc
If $(U,\SREV,\TREV)$ is an irreducible triplet of ${\mathcal I}^{-+}_0$ with $U^+$ irreducible, then $m\in\ZZ$ exists such that
${\mathcal H}=L_2(\RR^3,d\nu)\oplus L_2(\RR^3,d\nu)$ and
\par\noindent
$P_0=\left[\begin{array}{cc}p_0&0\cr 0&-p_0\end{array}\right]$,\quad $P_j=
\left[\begin{array}{cc}p_j&0\cr 0&p_j\end{array}\right]$,
\par\noindent
$J_j=\left[\begin{array}{cc}{\textsf J}_j^{(0)}+\textsf j_j&0\cr 0&{\textsf J}_j^{(0)}-\textsf j_j\end{array}\right]$,\quad
$K_j=\left[\begin{array}{cc}{\textsf K}_j^{(0)}+\textsf k_j&0\cr 0&-{\textsf K}_j^{(0)}+\textsf k_j\end{array}\right]$,
\par\noindent
where\quad
$\textsf j_1=\frac{m}{2}\frac{p_1p_0}{p_1^2+p_2^2}$,\quad
$\textsf j_2=\frac{m}{2}\frac{p_2p_0}{p_1^2+p_2^2}$,\quad
$\textsf j_3=0$,\quad
$\textsf k_1=-\frac{m}{2}\frac{p_2p_3}{p_1^2+p_2^2}$,\quad
$\textsf k_2=\frac{m}{2}\frac{p_3p_1}{p_1^2+p_2^2}$,\quad
$\textsf k_3=0$.
\vskip.5pc\noindent
With $m=0$ there are six triplets, each characterized by a different pair $(\TREV_n,\SREV_n)$, $n=1,2,...,6$.
\vskip.5pc\noindent
For every $m\neq 0$ in ${\mathcal I}^{-+}_0$ there are two triplets with different pairs  $(\TREV_a,\SREV_a)$
and $(\TREV_b,\SREV_b)$.}
\vskip.5pc
{\bf Remark 4.1.}
{\sl For the zero mass case
the {\sl helicity} operator $\hat\lambda=\frac{{\bf J}\cdot{\bf P}}{p_0}$ plays an important role.
Theorem 4.1 and (5) imply \cite{Nistico2} that $\hat\lambda=0$ for the triplets in 
${\mathcal I}_0^\pm$. 
\par
Using theorem 4.1 we see that $\hat\lambda=-\frac{m}{2}$ 
for every triplet of ${\mathcal I}_0^{-+}$
with $U^+$ irreducible. }
\subsection{The theories of elementary free particle with zero mass}
The possible theories of elementary free particle with zero mass can now be identified by 
selecting the triplets that admit a unique position operator.
One conclusion shared by the past approaches states 
that no position operator exists for massless particles with non-zero helicity.
Yet, the theoretical structures where such non-existence is proven \cite{Weinberg1}\cite{Jordan1}
are triplets where $\SREV$ is unitary and $\TREV$ is anti-unitary.
The present approach highlights that
this is a serious shortcoming, because according to theorem 3.1 these structures must be triplets in 
${\mathcal I}_0^+$ or ${\mathcal I}_0^-$.
But according to section 4.1 irreducible triplets with non-zero helicity
can exist only in ${\mathcal I}_0^{-+}$. Therefore, these proofs do not apply.
\par
In fact our approach proves the following theorems \cite{Nistico2}.
\vskip.5pc
{\bf Theorem 4.2.}
{\sl 
If $\hat\lambda\neq 0$, then in every triplet of ${\mathcal I}_0$
there is no three-operator satisfying $({\it Q}.1)$ and (4.i), (4.ii).}
\vskip.8pc
{\bf Theorem 4.3.}
{\sl For the triplet of $\;{\mathcal I}_0^+$ or of $\;{\mathcal I}_0^-$, with $U$ irreducible, there is only one three-operator
satisfying $({\it Q}.1)$ and $(4)$, namely Newton-Wigner operator 
${\bf Q}={\bf F}$.}
\vskip.5pc
Since the search for a position operator must be restricted to triplets with $\hat\lambda=0$, in ${\mathcal I}_0^{-+}$
only triplets with $m=0$ have to be checked.
\vskip.5pc
{\bf Theorem 4.4. }{\sl
The triplets of ${\mathcal I}_0^{-+}$ with a three-operator satisfying $({\it Q}.1)$ and $(4)$
are three of the six triplets with $m=0$ of theorem 4.1, characterized by 
$\TREV_1=\left[\begin{array}{cc}0&1\cr 1&0\end{array}\right]$,
$\SREV_1={\mathcal K}\left[\begin{array}{cc}0&1\cr1&0\end{array}\right]$, by 
$\TREV_2=\TREV_1$,$\SREV_2=\Upsilon\left[\begin{array}{cc}1& 0\cr0&-1\end{array}\right]$
and by 
$\TREV_3={\mathcal K\Upsilon}\left[\begin{array}{cc}0&1\\ 1&0\end{array}\right]$;\quad
$\SREV_3=\left[\begin{array}{cc}0&1\\- 1&0\end{array}\right]{\mathcal K}$.
\par\noindent
In all the three theories ${\bf Q}=\left[\begin{array}{cc}{\bf F}&0\cr 0&{\bf F}\end{array}\right]$.
}
\subsection{Conclusions for the zero mass case}
The current literature in fact restricts the search for theories of massless elementary free particle 
to triplets with $\TREV$ anti-unitary and $\SREV$ unitary, i.e. to triplets of ${\mathcal I}_0^+$ and
${\mathcal I}_0^-$. Our approach proves that consistent theories can be developed also if $\TREV$ is unitary or $\SREV$ 
is anti-unitary.
As a consequence, the class of possible theories extends to include a subclass of ${\mathcal I}_0^{-+}$.
\par
Furthermore, the non-existence proofs of a position operator for non zero helicity massless particles extends to
the larger class of possible theories, because the operators $\TREV$ and $\SREV$ play no role in the new theorem 4.2.


\bibliography{SciPost_Example_BiBTeX_File.bib}

\nolinenumbers

\end{document}